\documentclass[%        	Class options:
aps,%                   	American Physical Society
prc,%                   	Physical Review D
showpacs,%              	Displays PACS after abstract
superscriptaddress,%    	Authors' addresses linked with superscripts
nofootinbib,%           	Does not treat footnotes as references
floatfix]%              	Fixes float errors
{revtex4}%              	REVTEX 4 Package used
\usepackage{graphicx,%  	Default Latex 2eps package for embedding figures
longtable}%             	Useful for long table
\begin{document}
%======================================================================================
\title{Addendum to:
Quasiparticle random phase approximation uncertainties\\ and their correlations in the analysis of $0\nu\beta\beta$ decay}
%--------------------------------------------------------------------------------------
%
\author{        Amand~Faessler}
\affiliation{   Institute of Theoretical Physics,
				University of Tuebingen, %\\
               72076 Tuebingen, Germany}
\author{        G.L.~Fogli}
\affiliation{   Istituto Nazionale di Fisica Nucleare, Sezione di Bari, %\\
               Via Orabona 4, 70126 Bari, Italy}
\author{        E.~Lisi}
\affiliation{   Istituto Nazionale di Fisica Nucleare, Sezione di Bari, %\\
               Via Orabona 4, 70126 Bari, Italy}
\author{        V.~Rodin} 
\thanks{ Now at Stuttgart Technology Center, Sony Deutschland GmbH, Hedelfinger Strasse 61, 
D-70327 Stuttgart, Germany.}
\affiliation{   Institute of Theoretical Physics,
				University of Tuebingen, %\\
               72076 Tuebingen, Germany}
\author{        A.M.~Rotunno}
\affiliation{   Dipartimento Interateneo di Fisica ``Michelangelo Merlin,'' %\\
               Via Amendola 173, 70126 Bari, Italy}%
\author{        F.~\v{S}imkovic}
\affiliation{	Department of Nuclear Physics and Biophysics,
				Comenius University, %\\
				Mlynsk\'a dolina F1, SK--842 15 Bratislava, Slovakia}
\affiliation{	Bogoliubov Laboratory of Theoretical Physics, JINR, %\\
				141980 Dubna, Moscow Region, Russia}
\begin{abstract}%.......................................................................
\vspace*{1cm}
In a previous article [Phys.\ Rev.\ D {\bf 79}, 053001 (2009)] we estimated the 
correlated uncertainties associated to the nuclear matrix elements (NME) of neutrinoless double beta
decay ($0\nu\beta\beta$) within the quasiparticle random phase approximation 
(QRPA). Such estimates encompass recent independent calculations of NMEs, and can thus
still provide a fair representation of the nuclear model uncertainties. 
In this context, we compare the claim of $0\nu\beta\beta$ decay 
in $^{76}$Ge with recent negative results in $^{136}$Xe and in other nuclei, 
and we infer the lifetime ranges allowed or excluded at 90\% C.L.
We also highlight some issues that should be addressed in order to properly 
compare and combine results coming from different $0\nu\beta\beta$ candidate nuclei. 
\end{abstract}%.........................................................................
\medskip
\pacs{%PACS Numbers:
23.40.-s, 21.60.Jz, 02.50.-r} 
\maketitle

%\vspace*{0.2cm}
%%%%%%%%%%%%%%%%%%%%%%%%%%%%%%%%%%%%%%%%%%%%%%%%%%%%%%%%%%%%%%%%%%%%%%%%%%%%
%%%%%%%%%%%%%%%%%%%%%%%%%%%%%%%%%%%%%%%%%%%%%%%%%%%%%%%%%%%%%%%%%%%%%%%%%%%%
\section{Introduction \label{SecI}}
%%%%%%%%%%%%%%%%%%%%%%%%%%%%%%%%%%%%%%%%%%%%%%%%%%%%%%%%%%%%%%%%%%%%%%%%%%%%
%%%%%%%%%%%%%%%%%%%%%%%%%%%%%%%%%%%%%%%%%%%%%%%%%%%%%%%%%%%%%%%%%%%%%%%%%%%%

In a previous paper \cite{Ours} we presented the results of a systematic evaluation of nuclear matrix elements (NME)
and of their correlated uncertainties for the neutrinoless double beta decay process ($0\nu\beta\beta$) in
different nuclei, within the quasiparticle random phase approximation (QRPA) and the standard framework
of light Majorana neutrinos with effective mass $m_{\beta\beta}$. In particular, in \cite{Ours}
we discussed in the joint statistical distribution of the NME values $|M'_i|$ which govern,
together with the phase space $G_i$, the
decay half life $T_i$ in the $i$-th candidate nucleus,
%...................................................
\begin{equation}
\label{Ti}
T_i^{-1} = G_i\, |M'_i|^2\, m^2_{\beta\beta}\ ,
\end{equation}
%...................................................
with $i$ spanning the set
\begin{equation}
i=
{}^{76}\mathrm{Ge},\ 
{}^{82}\mathrm{Se},\ 
{}^{96}\mathrm{Zr},\ 
{}^{100}\mathrm{Mo},\  
{}^{116}\mathrm{Cd},\ 
{}^{128}\mathrm{Te},\ 
{}^{130}\mathrm{Te},\ 
{}^{136}\mathrm{Xe}\ . 
\end{equation}

We emphasized that the correlations among the NME uncertainties are sizable and play a relevant role
in comparing $0\nu\beta\beta$ data from different nuclei, including the $^{76}$Ge decay events
claimed by Klapdor {\em et al.} in \cite{Kl04,Kl06}. 

Recently, important new limits  on the  $^{136}$Xe half life have been obtained
by the experiments EXO-200 \cite{EXO0} and KamLAND-Zen \cite{KLZ0}, which have reached, for various
choices of NME calculations, a 90\% C.L.\ sensitivity to $m_{\beta\beta}$ largely overlapping 
with the $m_{\beta\beta}$ range favored by the $^{76}$Ge claim \cite{EXO0,KLZ0}. 
These advances have prompted us to use the results in \cite{Ours} 
to perform a systematic comparison of $^{136}$Xe half-life limits with those obtained in
other nuclei and with the $^{76}$Ge claim. The comparison is worked out in detail in the next Section, 
in terms of half-life ranges allowed or excluded at 90\% C.L. Once more, NME covariances are shown to 
play a relevant role in the $0\nu\beta\beta$ phenomenology. We also point out that, in order to fully exploit
the implications of upcoming $0\nu\beta\beta$ results, it is generally advisable to discuss in some detail
the probability distributions of both the experimental half lives and the theoretical NMEs.

We remark that, in this Addendum to \cite{Ours}, we adopt the same
NME and covariances as computed therein. 
A systematic update of \cite{Ours} would require extensive QRPA calculations of hundreds of 
NME, which are left to a future study. However, as we argue in the Appendix, the uncertainties
in \cite{Ours} are conservative enough to embrace recent NME calculations, using either the QRPA
or independent theoretical frameworks. We conclude that the NME estimates in \cite{Ours} still provide
a fair representation of the current spread of theoretical $0\nu\beta\beta$  calculations.  

\section{Comparison of the $0\nu\beta\beta$ claim with recent results}

In this section we briefly review, for the sake of completeness, 
the notation and conventions used in \cite{Ours} and
the implications of the $0\nu\beta\beta$ claim in ${}^{76}$Ge for different nuclei.
Then we compare such implications with recent experimental data, most notably from ${}^{136}$Xe in EXO-200 and KamLAND-Zen,
and discuss the regions allowed or excluded at $90\%$ C.L. 
We remind that gaussian
uncertainties at $n$ standard deviations on a given parameter correspond to projections of $\Delta \chi^2=n^2$ regions
on that parameter, and
that $90\%$ C.L.\ uncertainties correspond to $n=1.64$.

\subsection{Notation, conventions, and implications of ${}^{76}$Ge  $0\nu\beta\beta$ claim}
 
As in \cite{Ours}, we linearize Eq.~(\ref{Ti}) as
%...................................................
\begin{equation}
\label{linear}
\tau_i = \gamma_i - 2\eta_i - 2\mu\ ,
\end{equation}
%...................................................
by taking logarithms of the relevant $0\nu\beta\beta$ quantities in appropriate units:
%...........................................................................
\begin{eqnarray}
\label{taui}
\tau_i   &=&   \log_{10}(T_i/\mathrm{y})\ ,\\
\label{gammai}
-\gamma_i &=&   \log_{10} [G_i/(\mathrm{y}^{-1}\mathrm{eV}^{-2})]\ ,\\
\label{etai}
\eta_i   &=&   \log_{10} |M'_i|\ ,\\
\label{mu}
\mu      &=&   \log_{10} (m_{\beta\beta}/\mathrm{eV})\ .
\end{eqnarray}
%...........................................................................

The NME central values with their one-standard-deviation errors are denoted as 
%...........................................................................
\begin{equation}
\label{Eetai}
\eta_i  =   \eta_i^0\pm \sigma_i\ ,
\end{equation}
%...........................................................................
where the $\sigma_i$ are positively correlated
through a matrix $\rho_{ij}$. 
Table~I in \cite{Ours} reports the numerical values of
$\gamma_i$, $\eta_i^0$, $\sigma_i$, and $\rho_{ij}$, which are adopted hereafter.

Let us assume that $0\nu\beta\beta$ decay has been experimentally observed in 
$i={}^{76}$Ge as claimed by Klapdor {\em et al.} \cite{Kl04,Kl06}, with (logarithmic) half life given 
at $\pm1\sigma$ as \cite{Ours}:   
%...........................................................................
\begin{eqnarray}
\tau_i   &=&   \tau_i^0\pm s_i\\ 
         &=& 25.355 \pm 0.072\ (i={}^{76}\mathrm{Ge})\ .
\label{tauge}
\end{eqnarray}
%...........................................................................
Then, Eq.~(\ref{linear}) predicts the following half life in a different nucleus $j\neq i$ \cite{Ours}
%........................
\begin{equation}
\tau_j = \tau_j^0 \pm s_j\ ,
\end{equation}
%..........................
where
%............................
\begin{equation}
\label{tauj}
\tau_j^0 = \tau_i^0 + (\gamma_j-\gamma_i) - 2(\eta^0_j-\eta^0_i)\ ,
\end{equation}
%........................
and
%.....................
\begin{equation}
\label{sj}
s_j^2 = s^2_i + 4(\sigma^2_i+\sigma^2_j-2\rho_{ij}\sigma_i\sigma_j) \ , 
\end{equation}
%................................................................
the $(s_i,\,s_j)$ correlation being given by \cite{Ours}
%..................................
\begin{equation}
\label{rij}
r_{ij} = \frac{s_i}{s_j}\ (i\neq j)\ .
\end{equation}
%...................................
We add here that, for two nuclei $j$ and $k$ different from $i={}^{76}$Ge, the
$(s_j,\,s_k)$ correlation is given by
%..................................
\begin{equation}
\label{rjk}
r_{jk} = \frac{s^2_j+s^2_k-4(\sigma^2_j+\sigma^2_k-2\rho_{jk}\sigma_j\sigma_k)}{ 2s_j s_k}\ (j\neq i\neq k)\ ,
\end{equation}
%...................................
see also the Appendix.

\subsection{Applications and comparison with recent data}

Except for the claim in \cite{Kl04,Kl06}, all other $0\nu\beta\beta$ experiments report negative results to date. 
Table~\ref{bestlimits} shows the current best limits at 90\% C.L.\ on the $0\nu\beta\beta$ half life in different nuclei $j$
(i.e., $T^{90}_j$ and its logarithm $\tau^{90}_j$). Particularly important are the recent limits on $^{136}$Xe coming from 
EXO-200 ($T^{90}=1.6\times 10^{25}$~y) \cite{EXO0} and from KamLAND-Zen ($T^{90}=1.9\times 10^{25}$~y). A statistical combination of the negative results in \cite{EXO0,KLZ0}
is attempted in Ref.~\cite{KLZ0}, where the combined ``EXO~$\oplus$~KL-Zen'' limit $T^{90}({}^{136}\mathrm{Xe})=3.4\times 10^{25}$~y is quoted, 
as reported in Table~\ref{bestlimits} and adopted hereafter.  

%==============================================================================================================
\begin{table}[t]
\caption{Best current limits on half-lives at 90\% C.L. ($T_j>T_j^{90}$ and $\tau_j>\tau_j^{90}$) for
different nuclei $j$. \label{bestlimits}}
\begin{ruledtabular}
\begin{tabular}{ccccc}
$j$ & $T_j^{90}/\mathrm{y}$ & $\tau_j^{90}$ & Experiment & Ref.  \\[1mm]
\hline
$^{76}$Ge&  $1.6\times10^{25}$ & 25.204 & IGEX       & \protect\cite{IGEX} \\
$^{82}$Se&  $3.6\times10^{23}$ & 23.556 & NEMO-3     & \protect\cite{Ba12} \\
$^{96}$Zr&  $9.2\times10^{21}$ & 21.964 & NEMO-3     & \protect\cite{Ba12} \\
$^{100}$Mo& $1.1\times10^{24}$ & 24.041 & NEMO-3     & \protect\cite{Ba12} \\
$^{116}$Cd& $1.7\times10^{23}$ & 23.230 & Solotvina  & \protect\cite{Solo} \\
$^{128}$Te& $7.7\times10^{24}$ & 24.886 & Geochem.   & \protect\cite{Geoc} \\
$^{130}$Te& $2.8\times10^{24}$ & 24.447 & CUORICINO  & \protect\cite{Cuor} \\
$^{136}$Xe& $3.4\times10^{25}$ & 25.531 & EXO~$\oplus$~KL-Zen       & \protect\cite{EXO0,KLZ0} \\
\end{tabular}
\end{ruledtabular}
\end{table}
%==============================================================================================================

Figure~1 shows the limits reported in Table~\ref{bestlimits} (one-sided bands), together with the 90\%
C.L.\ ranges implied by the ${}^{76}$Ge claim, as derived from Eqs.~(\ref{tauge})--(\ref{sj}) with 
errors inflated by $\times 1.64$. It can be seen that, for the first time, there is a significant overlap 
between the half-life limit in one nucleus ($^{136}$Xe) and the corresponding region favored by Klapdor's claim 
for a given set of NME, as also emphasized in the experimental papers \cite{EXO0,KLZ0}. This situation
should be contrasted with the analogous Fig.~5 in \cite{Ours}, where no overlap emerged.

In principle, the next logical step should be a combination of positive and negative $0\nu\beta\beta$ results---within 
the adopted set of NME and their covariances---in order to 
evaluate the statistical consistency of the data (test of hypothesis) and to 
identify the range of $m_{\beta\beta}$ consistent with all the results 
(parameter estimation). This task would require the 
detailed knowledge of the probability distribution functions, not only for the NME (as attempted 
in \cite{Ours}), but also for the $^{76}$Ge and $^{136}$Xe half lives. However, the half-life likelihoods  
have not been published in the original papers 
\cite{Kl04,Kl06} and \cite{EXO0,KLZ0}. Reproducing or simulating the original data analyses is difficult, 
especially for the claimed signal in $^{76}$Ge, which involved a dedicated pulse-shape discrimination
\cite{Kl06}. It would be desirable that future $0\nu\beta\beta$ results are published also in terms
of likelihood or $\chi^2$ functions of the half-life, and not only in terms of specific  bounds, say, at 90\% C.L. 
Concerning theoretical uncertainties, we also note that after \cite{Ours} there has been no other independent
study of NME error correlations from the viewpoint of different nuclear models.
Therefore, we think that the conditions for a quantitative combination or a ``global fit'' of 
positive and negative $0\nu\beta\beta$ results are currently not warranted. 

%---------------------------------------------------------------------------
\begin{figure}[b]
%\vspace*{+1.0cm}
\hspace*{-1cm}
\includegraphics[scale=0.48]{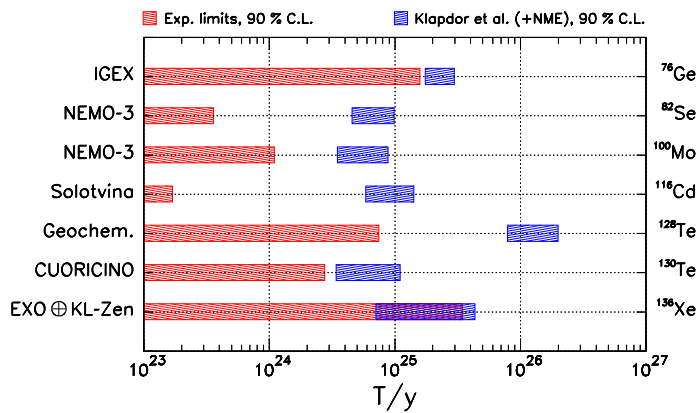}
%\vspace*{0.0cm}
\caption{\label{f01}
Range of half lives $T_i$ preferred at 90\% C.L. by the $0\nu\beta\beta$ claim
of \protect\cite{Kl06}, compared with the 90\% exclusion limits placed by other experiments. The 
comparison involves the NME and their errors, as well as their correlations. Note the
overlap of favored and disfavored ranges for $^{136}$Xe. This figure updates
Fig.~5 of \protect\cite{Ours}.
}
\end{figure}
%---------------------------------------------------------------------------
\newpage

%---------------------------------------------------------------------------
\begin{figure}[t]
%\vspace*{+1.0cm}
\hspace*{+1cm}
\includegraphics[scale=0.48]{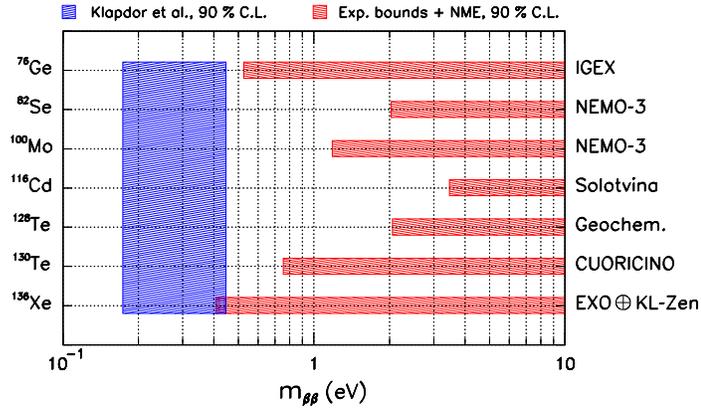}
\vspace*{-0.1cm}
\caption{ \label{f02} Range of $m_{\beta\beta}$ allowed by the $0\nu\beta\beta$ 
claim of \protect\cite{Kl06}, compared with the limits placed by other experiments (all at 90\% C.L.). This
figure updates Fig.~3 of \protect\cite{Ours}.}
\end{figure}
%---------------------------------------------------------------------------

The above issues emerge, e.g., when one tries to translate the numbers in Table~\ref{bestlimits} and
Eq.~(\ref{tauge}) in terms of 90\% C.L.\ limits on $m_{\beta\beta}$ via Eq.~(\ref{linear}). In the
absence of the experimental likelihood functions for the half lives, the combination of one-sided 
experimental limits ($\tau^{90}$) with two-sided theoretical errors ($\pm 1.64\sigma_i$) is not obvious. 
Conservatively, one may combine linearly the experimental and theoretical ranges at $90\%$ C.L.\ as proposed in \cite{Ours},
at the price of loosing statistical power. Figure~2 shows the results of such combination, in terms of
favored and disfavored ranges of $m_{\beta\beta}$. It can be noticed that
the $^{136}$Xe limits overlap with the range favored by the $^{76}$Ge claim, but
not as much as in Fig.~1, signaling the loss of statistical information. Therefore, we prefer to show
the following results directly 
in terms of the observable half lives $T_i$ and not via $m_{\beta\beta}$.

Figure~3 shows the application of Eqs.~(\ref{tauge})--(\ref{rij}) in the plane charted 
by the half lives $(T_i,\,T_j)$ for $i={}^{76}$Ge and  $j={}^{136}$Xe, at 90\% C.L.
The horizontal band corresponds to Klapdor's claim, while the slanted band represents the  
theoretical range at the $\pm 1.64\sigma$ level \cite{Ours}.
Their combination provides the allowed ellipse, which is however largely disfavored by the one-sided limit
placed by EXO~$\oplus$~KL-Zen (vertical bound). The surviving ellipse segment 
corresponds to
to  $T({}^{76}\mathrm{Ge})\simeq   2.0$--$2.9\times 10^{25}$~y and  
$T({}^{136}\mathrm{Xe})\simeq 3.4$--$4.3\times 10^{25}$~y. The upper ends of such
ranges set the limits required to test Klapdor's claim at 90\% C.L.  
Such an exercise could be repeated at higher C.L., if the corresponding  experimental limits
or the half-life likelihood were also published; this will become important in the future, 
since the $\sim 6\sigma$ signal claimed by Klapdor {\em et al.} 
\cite{Kl04,Kl06} should be tested at C.L.\ 
 definitely higher than 90\%.

%---------------------------------------------------------------------------
\begin{figure}[b]
%\vspace*{+1.0cm}
\hspace*{-0.6cm}
\includegraphics[scale=0.48]{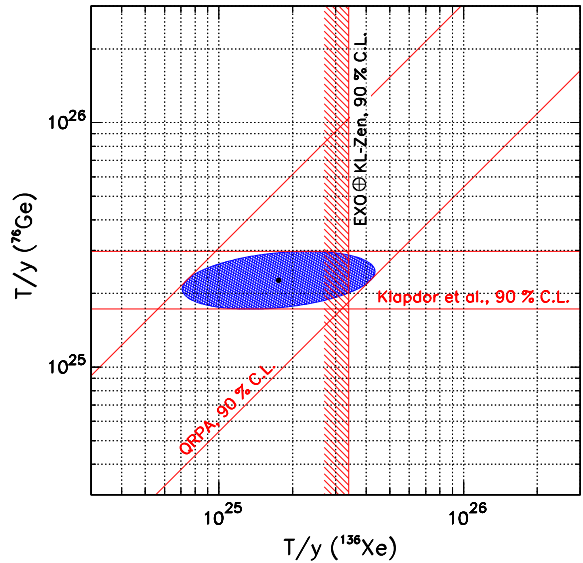}
\vspace*{-.2cm}
\caption{ \label{f03} 
Theoretical and experimental constraints in the plane charted
by the $0\nu\beta\beta$ half-lives of $^{76}$Ge and $^{136}$Xe.  Horizontal
band: range preferred by the  $0\nu\beta\beta$ claim of \protect\cite{Kl06}. Slanted band: constraint
placed by our QRPA estimates \protect\cite{Ours}. The combination 
provides the shaded ellipse, whose projection on the abscissa gives the range preferred at
90\% C.L. for the $^{136}$Xe half life. This range is largely disfavored
by the combined EXO~$\oplus$~KL-Zen results \protect\cite{EXO0,KLZ0} (vertical one-sided limit).}
\end{figure}
\newpage
%---------------------------------------------------------------------------

%---------------------------------------------------------------------------
\begin{figure}[t]
%\vspace*{+1.0cm}
%\hspace*{0cm}
\includegraphics[scale=0.48]{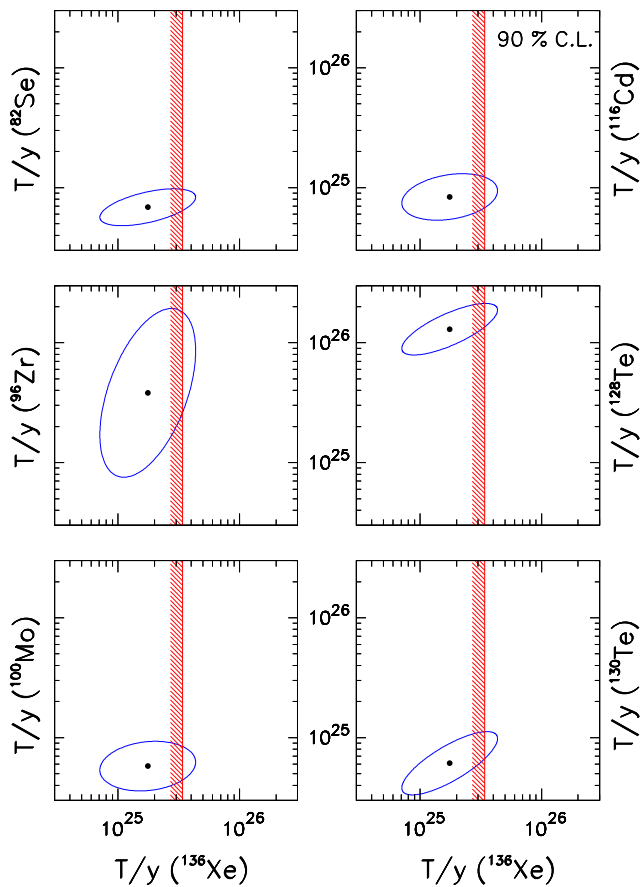}
\vspace*{-.1cm}
\caption{ \label{f04} 
Allowed regions (ellipses) as derived from Klapdor's claim \cite{Kl06} and the NME of \protect\cite{Ours},
in the plane charted by the half lives of $^{136}$Xe  and each  of the six nuclei 
${}^{82}\mathrm{Se}$, 
${}^{96}\mathrm{Zr}$, 
${}^{100}\mathrm{Mo}$,  
${}^{116}\mathrm{Cd}$, 
${}^{128}\mathrm{Te}$, and
${}^{130}\mathrm{Te}$. 
 A large fraction of each ellipse is excluded
by the combined EXO~$\oplus$~KL-Zen results \protect\cite{EXO0,KLZ0} (vertical one-sided limit).
All bounds are at 90\% C.L.\ on one variable.}
\end{figure}
%---------------------------------------------------------------------------

Figure~4 shows the application of Eqs.~(\ref{sj}) and (\ref{rjk}) in the six planes charted by the
the half lives $(T_j,\,T_k)$ for  $j={}^{136}$Xe and $k\neq{}^{136}$Xe,~${}^{76}$Ge. In each
panel, the ellipse represents the region favored by Klapdor's claim within the adopted
NME and their covariances, to be compared with the region excluded by
EXO~$\oplus$~KL-Zen at the same C.L.\ (vertical bound). As a consequence of the positive
correlation, the ellipse segment not excluded by the $^{136}$Xe limit corresponds, for 
each $k$-th nucleus,
to the higher end of the half-life range at 90\% C.L.\ For instance, in the
case of $^{130}$Te (lower right panel in Fig.~4), the current limit should be pushed from $2.8\times 10^{24}$~y up to
$\sim 0.7$--$1.1\times 10^{25}$~y, in order to cover the ellipse segment left out  
by the current EXO~$\oplus$~KL-Zen bound. If correlations
were neglected, the ellipse would not be tilted, and this requirement would (incorrectly) become less stringent.

We remark that the results shown in this section are based on the same
NME and covariances as in \cite{Ours}. In general, it would be useful to extend the theoretical
covariance analysis in further directions, including: ($i$) updated and improved QRPA calculations;
($ii$) other candidate $0\nu\beta\beta$ nuclei not considered in \cite{Ours}; $(iii)$ theoretical NME   
approaches different from the QRPA (see also the Appendix); $(iv)$ nonstandard decay mechanisms.
From the experimental viewpoint, we have emphasized the importance of publishing 
the probability distribution of the half life for each nucleus. All these refinements will become 
increasingly important in the next few years, since  a number of  
$0\nu\beta\beta$ experiments will provide highly significant data,
which must be eventually combined in proper theoretical and statistical frameworks.

\section{Summary}

In the previous work \cite{Ours} we presented estimates of NME and their covariances
for a set of candidate $0\nu\beta\beta$ nuclei, within the QRPA theoretical framework.
Such estimates still provide a fair representation of the spread of NME calculations
(see the Appendix). In this context, we have compared herein the claimed
$^{76}$Ge signal from \cite{Kl04,Kl06} with negative results from other experiments,
including the recent $^{136}$Xe limits placed by EXO-200 \cite{EXO0} and KamLAND-Zen \cite{KLZ0}. 
We have worked out favored and disfavored ranges at 90\% C.L.\ for each nucleus
and for couples of nuclei, in terms of either half lives $T_j$ (Figs.~1, 3 and 4) or of $m_{\beta\beta}$ 
(Fig.~2). In particular, we find that, in order to close the region currently allowed at 90\% C.L.\
by the $^{76}$Ge claim and by the $^{136}$Xe limit, one should cover 
either the range $T({}^{76}\mathrm{Ge})\simeq 2.0$--$2.9\times 10^{25}$~y or the range
$T({}^{136}\mathrm{Xe})\simeq 3.4$--$4.3\times 10^{25}$~y; alternatively, using a third
nucleus such as $^{130}$Te, one should cover the range  
$T({}^{130}\mathrm{Te})\simeq 0.7$--$1.1\times 10^{25}$~y (see also Fig.~4 for other nuclei). 
We remark that the theoretical NME covariances play a 
relevant role in these or similar estimates: their study should thus be further pursued,
not only within the QRPA \cite{Ours}, but also within other approaches as well as 
for nonstandard $0\nu\beta\beta$ mechanisms. We have also emphasized that
 experimental results should be given in terms of likelihood functions for the
decay half-life (rather than in terms of bounds at a fixed C.L.), in order to 
allow a proper combination of the experimental and theoretical uncertainties, 
which are equally important to derive constraints on 
$0\nu\beta\beta$ parameters.

\medskip

%%%%%%%%%%%%%%%%%%%%%%%%%%%%%%%%%%%%%%%%%%%%%%%%%%%%%%%%%%%%%%%%%%%%%%%%%%%%%%%%%%%%%%%%%%%%%%%%%%%%
%%%%%%%%%%%%%%%%%%%%%%%%%%%%%%%%%%%%%%%%%%%%%%%%%%%%%%%%%%%%%%%%%%%%%%%%%%%%%%%%%%%%%%%%%%%%%%%%%%%%

%\acknowledgments

{\em Acknowledgments.} 
The work of G.L.F., E.L., and A.M.R.\ is supported by the Italian Istituto Nazionale di Fisica 
Nucleare (INFN) and Ministero dell'Istruzione, dell'Universit\`a e della Ricerca 
(MIUR) through the ``Astroparticle Physics'' 
project. F.\v{S}.\ is supported by the VEGA Grant Agency
of the Slovak Republic under contract N.~1/0876/12.

\section*{APPENDIX}

%==============================================================================================================
\begin{table}[b]
\caption{Estimates of $\eta_i=\log_{10}|M'_i|$ for each nucleus, as derived from the
recent EDF \protect\cite{EDFS},  IBM-2 \protect\cite{IBM2}, and PHFB \protect\cite{PHFB}  calculations  
after appropriate rescaling, in order to match the conventions used in \cite{Ours}. 
The estimates of  \protect\cite{PHFB} 
refer only to a subset of nuclei. Also shown are the $\eta_i$ central values for two widely different
(R)QRPA models recently reported in \protect\cite{Fa12}.  The corresponding values of
the adopted effective axial coupling $g_A$ are also reported.
The last two rows report the upper and lower
ends of our $3\sigma$ range $\eta_i^0\pm 3 \sigma_i$ as taken from \cite{Ours}, 
which largely encompass the above $\eta_i$ values.  
\label{Comp}}
\begin{ruledtabular}
\begin{tabular}{cccrrrrrrrr}
Ref.  & Model & $g_A$ & $^{76}$Ge & $^{82}$Se & $^{96}$Zr & $^{100}$Mo & $^{116}$Cd & 
$^{128}$Te & $^{130}$Te & $^{136}$Xe \\
\hline
\protect\cite{EDFS} & EDF   & $1.25\times 0.74$ & 0.617 &  0.577 &  0.707 &  0.663 &  0.629 &  0.570 &  0.665 &  0.571 \\
\protect\cite{IBM2} & IBM-2 & 1.269  			& 0.721 &  0.622 &  0.390 &  0.564 &  0.421 &  0.616 &  0.562 &  0.467 \\
\protect\cite{PHFB} & PHFB  & 1.254 			&       &        &  0.474 &  0.813 &        &  0.586 &  0.623 &        \\
\protect\cite{PHFB} & PHFB  & 1.0  				&       &        &  0.317 &  0.658 &        &  0.433 &  0.470 &        \\
\protect\cite{Fa12} & RQRPA (Jastrow) & 1.0 	& 0.535 &  0.460 &  0.045 &  0.365 &  0.289 &  0.401 &  0.291 &  0.160 \\
\protect\cite{Fa12} & QRPA (CD-Bonn) & 1.25 	& 0.797 &  0.748 &  0.316 &  0.717 &  0.597 &  0.736 &  0.689 &  0.468 \\
\hline
\protect\cite{Ours} & \multicolumn{2}{c}{Lower limit of our $3\sigma$ range }
                                    & 0.269 &  0.166 &$-0.703$&  0.017 &$-0.046$ &  0.072 &  0.024 &$-0.307$  \\
\protect\cite{Ours} &  \multicolumn{2}{c}{Upper limit of our $3\sigma$ range }
                                    & 1.001 &  0.976 &  0.779 &  0.989 & 0.854  &  0.996 &  0.972 &  0.815
\end{tabular}
\end{ruledtabular}
\end{table}
%==============================================================================================================

In the Appendix of Ref.~\cite{Ours}, we clarified the role of different conventions about the phase space
factors $G_i$ and the nuclear matrix elements $|M'_i|$. We also compared our QRPA estimates for the NME logarithms
$\eta_i$ with those calculated independently in \cite{Su08} (within the QRPA) and in \cite{She1,She2} 
(within the shell model, SM), which were all encompassed by our $\pm 3\sigma$ ranges ($\eta^0_i\pm 3\sigma_i$). In this Appendix,
we show that such ranges also embrace more recent NME calculations performed via the Energy Density Functional method
(EDF) \cite{EDFS}, the microscopic Interacting Boson Model (IBM-2) \cite{IBM2}, and the projected Hartree-Fock-Bogoliubov
method (PHFB) \protect\cite{PHFB}, as well as the Renormalized QRPA (RQRPA) approach \cite{Fa12}. The last
calculation is particularly relevant with respect to \cite{Ours}, since it
embeds the recent experimental determination of the $2\nu2\beta$ half life in $^{136}$Xe \cite{EXO2,KLZ2}
(not available at the time of \cite{Ours}), which fixes the so-called $g_{pp}$ parameter of the QRPA.

Different NME calculations may use slightly different phase space factors (see the recent detailed evaluation in \cite{PhSp}),
different values of the nuclear radius parameter $r_0$, and different conventions. In order to make a homogeneous
comparison with the $\eta$ values and the conventions of \cite{Ours}, we use the fact that the papers \cite{EDFS,IBM2,PHFB,Fa12}
contain tables of estimated half lives ($\tau_i$) at fixed values of the effective Majorana mass ($\mu$); then, by adopting
the same phase space factor ($\gamma_i$) reported in \cite{Ours}, the NME values $(\eta_i)$ can be calculated via Eq.~(\ref{linear})
and can be directly compared with those in \cite{Ours}. For this purpose, we use the half lives in Table~I of \cite{EDFS},
in  Table~III of \cite{IBM2}, in Table~IV of \cite{PHFB} (central values), and in Table~I of \cite{Fa12} (central
values). From \cite{Fa12} we select two options, RQRPA with Jastrow short-range correlations, and QRPA with
CD-Bonn potential, which generally provide the lowest and highest NME values, respectively \cite{Fa12}.
We note that the papers \cite{PHFB,Fa12} also report estimated NME uncertainties, but not their covariance matrix; 
in any case, the errors estimated in \cite{Ours} are generally more conservative than those in \cite{PHFB,Fa12}.

Table~\ref{Comp} reports the effective $\eta_i$ of such recent EDF, IBM-2, PHFB, and (R)QRPA calculations, to be 
compared with the $\pm 3\sigma$ ranges from \cite{Ours} in the last two rows. Such ranges largely embrace all the above values.
Actually, almost all the NME in Table~\ref{Comp} are contained in the 90\% C.L.\ ranges ($\eta_i^0\pm 1.64 \sigma_i$, not shown). 
We conclude that Table~I of \cite{Ours} still provide a reasonable and conservative evaluation of the NME 
$\eta_{i}$ and of their variances $\sigma_i^2=\mathrm{var}(\eta_i)$. 

Concerning the NME covariances, $\mathrm{cov}(\eta_i,\,\eta_j)=\rho_{ij}\sigma_i\sigma_j$, no comparison 
is possible within the current literature, since they have been evaluated only in \cite{Ours}. Here  
we just remind their crucial role, by deriving the last three equations in Sec.~II~A. 
Since the experimental error $s_i$ ($i={}^{76}$Ge) in Eq.~(\ref{tauge}) is independent from any theoretical error $\sigma_j$, it is $\mathrm{cov}(\tau_i,\,\eta_j)=0$; in addition,
phase space uncertainties are currently negligible, $\mathrm{var}(\gamma_j)\simeq 0$. 
Thus, in the nontrivial case $j\neq i$, propagation of errors in Eq.~(\ref{tauj}) gives
%..........................
\begin{eqnarray}
s^2_j &\equiv& \mathrm{var}(\tau_j) \nonumber \\ 
&=&\mathrm{var}(\tau_i)+4[\mathrm{var}(\eta_j)+\mathrm{var}(\eta_i)-2
\mathrm{cov}(\eta_i,\,\eta_j)] \nonumber \\
&=& s^2_i + 4 (\sigma^2_i + \sigma^2_j - 2\rho_{ij}\sigma_i\sigma_j)
\end{eqnarray}
%............................
and 
%..........................
\begin{eqnarray}
r_{ij}s_i s_j &\equiv & \mathrm{cov}(\tau_i,\,\tau_j)\nonumber \\
&=&\mathrm{cov}(\tau_i,\,\tau_i)\nonumber \\
&=& s^2_i\ ,
\end{eqnarray}
%............................
as reported in Eqs.~(\ref{sj}) and (\ref{rij}). Similarly, for $j\neq i$ and $k\neq i$ (where $i={}^{76}$Ge), it is 
%..........................
\begin{eqnarray}
r_{jk}s_j s_k &\equiv & \mathrm{cov}(\tau_j,\,\tau_k)\nonumber \\
&=&\mathrm{cov}(\tau_i,\,\tau_i)
   +4\mathrm{cov}(\eta_i,\,\eta_i)
   -4\mathrm{cov}(\eta_i,\,\eta_k)-4\mathrm{cov}(\eta_i,\,\eta_j)
   +4\mathrm{cov}(\eta_j,\,\eta_k) \nonumber \\
&=& s^2_i+4\sigma^2_i - 4\rho_{ij}\sigma_i\sigma_j - 4\rho_{ik}\sigma_i\sigma_k + 4\rho_{jk}\sigma_j\sigma_k
\nonumber\\
&=& \frac{1}{2}(s^2_j+s^2_k)-2(\sigma^2_j+\sigma^2_k-2\rho_{jk}\sigma_j\sigma_k)\ ,
\end{eqnarray}
%............................
as reported in Eq.~(\ref{rjk}). The relevance of NME covariances clearly emerges in the above equations.

%\newpage

%%%%%%%%%%%%%%%%%%%%%%%%%%%%%%%%%%%%%%%%%%%%%%%%%%%%%%%%%%%%%%%%%%%%%%%%%%%%%%%%%%%%%%%%%%%%%%%%

\end{document}